\documentclass[twoside,10pt]{article}

\usepackage[T1]{fontenc}
\usepackage[utf8]{inputenc}
\usepackage{lmodern}     % Latin Modern fonts

\usepackage[english]{babel}

\usepackage{hyperref}    % usually these two should be loaded, and "nohyperref" shouldn't
\usepackage{bookmark}    % usually these two should be loaded, and "nohyperref" shouldn't
\usepackage{xcolor,soulutf8}
\sethlcolor{yellow}

\usepackage[pdftex]{graphicx}

\usepackage[hmarginratio=1:1,margin=32mm]{geometry}

\usepackage[]{tocbibind}         % Adding LoF, LoT and ToC itself to ToC
\usepackage[longnamesfirst]{natbib}
\setcitestyle{numbers,square,comma}

\usepackage{lipsum}      % Write "Lorem ipsum..." placeholder text

\usepackage{paralist}
\usepackage{enumitem}
\usepackage{booktabs}
\usepackage{array}
\usepackage{tabularx}
\newcolumntype{Y}{>{\centering\arraybackslash}X}
\newcolumntype{Z}{>{\raggedleft\arraybackslash}X}
\usepackage{multirow}

\usepackage{wasysym}
\usepackage{amssymb}
\usepackage{amsmath}
\usepackage{mathtools}
\usepackage{nicefrac}
\usepackage{textcomp}

\usepackage{verbatim}
\usepackage{wrapfig}     % Wrap text around graphics (i.e. tip environment)

\usepackage{etoolbox}

\usepackage{relsize}

\usepackage{caption}
\usepackage{subcaption}

\usepackage[blocks]{authblk}
\usepackage[acronym,hyperfirst=false,style=index,nogroupskip]{glossaries}
\makeglossaries
\newacronym{lpres}{LPRES}{Liquid-Propellant Rocket Engine Simulation}
\newacronym{espss}{ESPSS}{European Space Propulsion System Simulation}
\newacronym{lprm}{LPRM}{liquid-propellant rocket motor}
\newacronym{ssme}{SSME}{Space Shuttle main engine}
\newacronym{dae}{DAE}{differential algebraic equation}
\newacronym{ode}{ODE}{ordinary differential equation}
\newacronym{el}{EL}{EcosimPro Language}
\newacronym{proosis}{PROOSIS\protect\supers{\textregistered}}{PRopulsion Object-Oriented SImulation Software}
\newacronym{esa}{ESA}{European Space Agency}
\newacronym{eclss}{ECLSS}{environmental control and life support system}
\newacronym{upm}{UPM}{Universidad Politécnica de Madrid}   % for the double-blind review, HIDE THIS !!!
\newacronym{jitt}{JiTT}{just-in-time teaching}
\newacronym{pbl}{PBL}{problem-based learning}
\newacronym{cscl}{CSCL}{computer-supported collaborative learning}   % we do NOT use this one, as of yet...

\usepackage[gen]{eurosym}
\DeclareUnicodeCharacter{20AC}{\euro{}}

\graphicspath
{
  {./main-images/}
}

\newcommand{\supers}[1]{\ensuremath{^{\textrm{#1}}}}

\usepackage{abstract} % Allows abstract customization
\usepackage{url}
\makeatletter
\def\url@leostyle{%
  \@ifundefined{selectfont}{\def\UrlFont{\sf}}{\def\UrlFont{\small\ttfamily}}}
\makeatother
\title{\vspace{-3em}\fontsize{20pt}{24pt}\selectfont\textbf{Developing a New Tool to Implement Computer-Supported Active Learning Strategies in the Engineering Classroom}\vspace{1em}} % Article title
\author[a]{Juan~M.~Tizón\thanks{E-mail:~\href{mailto:jm.tizon@upm.es}{\texttt{jm.tizon@upm.es}}---ORCID~iD:~\href{https://orcid.org/0000-0002-8687-6657}{\texttt{https://orcid.org/0000-0002-8687-6657}}}}
\author[a]{Pablo~Sierra\thanks{E-mail:~\href{mailto:pablo.sierra.heras@alumnos.upm.es}{\texttt{pablo.sierra.heras@alumnos.upm.es}}---ORCID~iD:~\href{https://orcid.org/0000-0001-7876-2724}{\texttt{https://orcid.org/0000-0001-7876-2724}}}}
\author[a]{Luis~Sánchez~de~León\thanks{Address all correspondence to this author: Luis Sánchez de León Peque, Departamento de Mecánica de Fluidos y Propulsión Aeroespacial, Escuela Técnica Superior de Ingeniería Aeronáutica y del Espacio, Universidad Politécnica de Madrid, Pza.~del Cardenal Cisneros~3, 28040--Madrid, Spain. E-mail:~\href{mailto:luis.sanchezdeleon@upm.es}{\texttt{luis.sanchezdeleon@upm.es}}---ORCID~iD:~\href{https://orcid.org/0000-0001-6815-3263}{\texttt{https://orcid.org/0000-0001-6815-3263}}}}     %\url{} works, but wrong size for the text in the footnote...!!!
\author[a]{Emilio~Navarro\thanks{E-mail:~\href{mailto:emilio.navarro@upm.es}{\texttt{emilio.navarro@upm.es}}}}
\author[b]{Javier~Vilá\thanks{E-mail:~\href{mailto:jvv@empre.es}{\texttt{jvv@empre.es}}}}
\author[b]{José~F.~Moral\thanks{E-mail:~\href{mailto:frj@empre.es}{\texttt{frj@empre.es}}}}
\affil[a]{\protect\vphantom{\Large B}Departamento de Mecánica de Fluidos y Propulsión Aeroespacial\\
    Escuela Técnica Superior de Ingeniería Aeronáutica y del Espacio (ETSIAE)\\
    Universidad Politécnica de Madrid (UPM)\\
    Pza.~del Cardenal Cisneros~3\\
    28040~Madrid, Spain}
\affil[b]{\protect\vphantom{\Large B}Empresarios Agrupados Internacional,~S.A.\\
    División de Simulación y Sistemas de Transporte\\
    c/~Magallanes~3\\
    28015~Madrid, Spain}
\date{}     % remove date: make the command 'blank'

\begin{document}
\maketitle

%%%%%%%%%%%%%%%%%%%%%%%%%%%%%%%%%%%%%%%%%%%%%%%%%%%%%%%%%%%%%%%%%%%%%%
%%%%%%%%%%%%%%%%%%%%%%%%%%%%%%%%%%%%%%%%%%%%%%%%%%%%%%%%%%%%%%%%%%%%%%
%%%%%%%%%%%%%%%%%%%%%%%%%%%%%%%%%%%%%%%%%%%%%%%%%%%%%%%%%%%%%%%%%%%%%%

\vspace{-2em}

\begin{abstract}
%%%% NON-STRUCTURED, "STANDARD" ABSTRACT !!!
\noindent Successful implementation of active learning strategies in the engineering classroom ---and in particular in certain subjects which are highly technological in nature such as, for instance, rocket engines and space propulsion--- means overcoming certain challenges that arise from the fact that these are extremely complex systems to analyze. In this paper, we address the specific means to overcome one of such challenges: the lack of readily available software tools that are suitable for implementing this sort of teaching strategies within the engineering training. In particular, we develop a new tool for the modeling and simulation of liquid-propellant rocket engines specially tailored for the classroom, taking a systematic approach to the development of such tool based on the needs of modern teaching practices. After a thorough review of the available literature on the topic, the few most critical features that our tool should have in order to serve its purported goal are identified. Subsequently, a pilot experience to assess the impact of the usage of said tool on the learners' performance was carried out, showcasing excellent results, both in terms of the students' perceived quality of their training as well as in terms of their grade of retention and understanding of the matter. The conclusions of this study, especially the guidelines for the development of software tools aimed at the classroom, nevertheless, should be applicable to any other highly technological discipline, extending the scope of this paper beyond merely the subject of rocket science in engineering.

\vphantom{B}

\noindent\textup{\textbf{Keywords}: Aerospace engineering; Educational software; Active learning; Simulation; Systems thinking.}
\end{abstract}

%%%%%%%%%%%%%%%%%%%%%%%%%%%%%%%%%%%%%%%%%%%%%%%%%%%%%%%%%%%%%%%%%%%%%%
%%%%%%%%%%%%%%%%%%%%%%%%%%%%%%%%%%%%%%%%%%%%%%%%%%%%%%%%%%%%%%%%%%%%%%
%%%%%%%%%%%%%%%%%%%%%%%%%%%%%%%%%%%%%%%%%%%%%%%%%%%%%%%%%%%%%%%%%%%%%%

\clearpage
\glsresetall   % re-explain acronyms as if they appeared for the first time
%\glsaddall
%\printglossary[type=\acronymtype,title=Abbreviations,toctitle=Abbreviations \&{} Nomenclature,nonumberlist]
\printglossary[type=\acronymtype,title=Abbreviations,nonumberlist]

%\begin{nomenclature}
%\entry{$\pi$}{total-to-total pressure ratio}
%\entry{$M$}{Mach number}
%\entry{$p$}{pressure (Pa)}
%\entry{$T$}{temperature (K)}
%\entry{$v$}{velocity (m/s)}

%\subsection*{Subscripts}
%\entry{${}_0$}{ambient/flight conditions}
%\entry{${}_t$}{total or stagnation magnitude}

%\end{nomenclature}

%%%%%%%%%%%%%%%%%%%%%%%%%%%%%%%%%%%%%%%%%%%%%%%%%%%%%%%%%%%%%%%%%%%%%%
%%%%%%%%%%%%%%%%%%%%%%%%%%%%%%%%%%%%%%%%%%%%%%%%%%%%%%%%%%%%%%%%%%%%%%
%%%%%%%%%%%%%%%%%%%%%%%%%%%%%%%%%%%%%%%%%%%%%%%%%%%%%%%%%%%%%%%%%%%%%%

\glsresetall   % re-explain acronyms as if they appeared for the first time
\section{Introduction}\label{sec:intro}

In the era of artificial intelligence and technological revolution that we are living today, higher education must evolve and adapt, preparing our graduate students to meet the demands of the working environment of tomorrow. As envisioned by Aoun \citep{robotProof}, the workers of tomorrow will compete in a much more flexible and demanding market, where jobs that are suitable to be automated will be done by robots, while humans will have to develop their creativity in order to remain competitive. In order for the higher education of today to be able to train the workers of tomorrow, the whole vision of the higher education itself must move towards new paradigms: training the students for a continuous, life-long, self-guided process of learning, to continually adapt and meet the ever changing needs of the market.

As such, and in particular within the engineering field, it will not suffice for the engineer of tomorrow to be an expert on a certain field or a specific part of a device, but rather a much more holistic view of the whole system will be needed. The fundamental understanding of the interactions between several disciplines, the merging between the technical knowledge of a particular device's functioning, data-analysis capabilities and technological ---i.e. computer software development--- expertise is what big companies in the engineering sector are after, even today \citep[p.~36]{robotProof}.

According to Aoun \citep[p.~xix]{robotProof}, the workers of tomorrow will need to develop new competencies, among which \emph{systems thinking} will be fundamental; and acquire new abilities, such as \emph{data literacy} and \emph{technological literacy}. For this purpose, Aoun defends the concepts of \emph{experiential learning} and \emph{cooperative education}, in which both the Industry as well as the University share the responsibility for the learners' training. However, the precise tools and means to achieve this sort of education, in a field such as engineering, is not irrevocably established yet. Among the many changes in teaching practices and the new paradigms in learning strategies that we have seen arising in the last decades, two aspects are common to the vast majority of them:
\begin{inparaenum}[\itshape a\upshape)]
    \item the use of technology and computers as a \emph{cognitive tool} \citep[as defined by][]{jonassen1996}; and
    \item the drive towards a more `active' role from the students in the classroom.
\end{inparaenum}

The challenges for implementing this sort of concepts in teaching practices within the engineering field, and in particular within the sector of rocket engines and space propulsion that we are dealing with, respond to some of its technical particularities: for starters, it is a field in which the complex interactions between the many components that form a single system (a rocket engine, in our case) cannot be understood without a profound knowledge of the `basic science' that underlies beneath, so that active learning is not an attainable paradigm without at least some time devoted to the traditional approach of the student absorbing knowledge from the lecturer in a more `passive' role; furthermore, the computational tools used in this sector for the simulation of rocket engines either fall within the category of in-house code developed by the manufacturers, and are consequently inaccessible for the students, or are just generic-purpose tools with such a complex modeling that require experts on that particular tool to be trained just to use them ---this is the case, for instance, of the \gls{espss} libraries developed for EcosimPro\supers{\textregistered}, that will be referred to in Section~\ref{sec:devel}---, which in turn would require a vast amount of time to be spent in such a training, hardly affordable within the time frame of a semester-long course.

In consequence, active learning and, in particular, inquiry-based methods have been used much less extensively in engineering than in many other sciences \citep{prince2006}. In view of the apparent lack of readily available software tools actually useful for the purposes of implementing active learning and \gls{pbl} paradigms in the engineering classroom ---more specifically, in teaching the subject of rocket engines---, the objectives of this project are:
\begin{inparaenum}[\itshape a\upshape)]
    \item to identify the major shortfalls of currently used software tools in the field for this purported goal;
    \item to systematically devise the features that a new tool should have in order to fulfill its expected usage; and
    \item to assess the impact of the usage of said developed tool on the learners' performance.
\end{inparaenum}
For this purpose, after the tool was developed, a pilot experience was carried out in the context of the Space Systems MSc program at the \gls{upm}.

%%%%%%%%%%%%%%%%%%%%%%%%%%%%%%%%%%%%%%%%%%%%%%%%%%%%%%%%%%%%%%%%%%%%%%
%%%%%%%%%%%%%%%%%%%%%%%%%%%%%%%%%%%%%%%%%%%%%%%%%%%%%%%%%%%%%%%%%%%%%%
%%%%%%%%%%%%%%%%%%%%%%%%%%%%%%%%%%%%%%%%%%%%%%%%%%%%%%%%%%%%%%%%%%%%%%

\section{Motivation}\label{sec:motivation}

The courses within any engineering MSc program are highly technological in nature. Basic knowledge of `science' has already been imparted in previous years and should be applied to solving problems in the real world. The trouble with such problems is often associated with two aspects: the degree of realism for describing the different elements, and the complexity ---understood as the presence of many interrelated components--- of the systems to analyze.

From the old paradigm of `passive' learning, in which students merely `absorb' knowledge from a lecture, a textbook, etc., teaching of such courses nowadays tries to place the students in a more active role. In the particular subject of \emph{Rocket Engines and Space Propulsion} that we are dealing with, nevertheless, there are several issues that somewhat limit the extent to which an active learning paradigm can be adopted. On the one hand, practical experience, hands-on, with an actual rocket engine in a laboratory is unaffordable, both in terms of cost and in terms of safety; on the other, classroom `activities', such as problem-solving carried out with ``pen and paper'', tend to be oversimplified (so that the exercises can actually be tackled ``by hand''), which triggers in the learners the perception of `disconnection' from reality: what they study feels meaningless, too simplistic and completely unrelated to the amazing complexity of the actual systems.

Computers, as \emph{cognitive tools}, therefore, stand as an interesting alternative. It must be noted, though, the key difference between `learning with' computers and `learning from' computers \citep{jonassen1996,campbell2010}. In `learning from', the student assumes a passive role, and uses the computer merely as a source of information (as they could get the information from listening to a lecture or reading a textbook); while in `learning with' computers, the emphasis is placed upon the active role of the learner: they must \emph{do} rather than \emph{watch}. In this regard, merely engaging the students in a few guided simulations will just not be enough: in the \emph{guided simulations} approach, the student merely uses a software tool to produce some data and \emph{watches} (following some guided steps from the instructor so that a certain software plots some results in the screen is not much different from just \emph{watching} the instructor plot those same results in a projector for the whole classroom). And hence there is a close link, and some mutual necessity, between using computers as a cognitive tool and implementing active inquiry-driven, self-guided learning; between `learning with' computers and \gls{pbl}, in which the learner \emph{does} actively engage in solving real-world, ``ill-structured'' problems.

According to \citep{campbell2010}, software tools can be used to alleviate the `cognitive burden' of solving huge systems of equations, so that the student focuses on gathering data and analyzing the results rather than on the mathematical apparatus needed to solve such problems; and thus to encourage scientific inquiry and self-guided learning. Although not every author agrees with the inquiry-driven learning paradigm \citep{settlage2007}, we share the view of many \citep{johnston2008,campbell2010} who consider \emph{inquiry} not as a mere \emph{tool} to learn something else, but as a scientific \emph{goal} in itself, worthwhile teaching in its own right.

Active learning strategies in engineering have been thoroughly discussed, among others, by Prince~et~al. in \citep{prince2004,prince2006}. In the \emph{Rocket Engines and Space Propulsion} subject, there is undoubtedly a need for at least some theoretical lectures, so that the fundamentals of these complex systems are explained, and hence pure \emph{discovery learning} without guidance from the instructor is not recommended \citep{prince2006}. Still, inquiry-driven learning shows generally good results \citep{prince2004}, and \gls{pbl}, at the very least, encourages better attitudes from the students \citep{prince2004}.   % De \citep{prince2004}, olvidarse de "collaborative" best vs. "individual", y de "cooperative" best vs. "competitive". A lo mejor citarlo en el Future Work???

In other scientific areas, highly dependent on data gathering to infer theories or draw conclusions, computer modeling has already proved itself as a valuable tool for increasing the students' understanding of the matter \citep[e.g.][]{carey2017}. Even from the mere fact of engaging students in modeling activities, other, more general, cognitive benefits ---such as the ability to scan data for patterns or make evidence-based predictions--- that are intrinsic to science and not specific of a particular field, are acquired by the students \citep{carey2017,schwarz2009,schwarz2005}.

In teaching physics, for instance, the usage of such modeling tools is widespread. We take from \citep{lopez2015} the subtle difference between \emph{modeling} and \emph{simulating}: in the former, the learner can modify the \emph{model} itself, according to the variables or effects that they think are relevant for the studied phenomena; while in the latter the learner can only modify the \emph{value} of the input variables, but is unable to define or change the \emph{relations} between those variables. This is a fundamental concept that has guided the development of our tool. In \citep{rodriguez2013}, the benefits of including computer modeling activities as a means to enhance the \gls{jitt} methodology is reported. Not only do the students get a more active role in the classroom by means of trying by themselves to answer the questions that the instructor poses in a mostly \emph{guided-inquiry} teaching process, but also the professor can identify the common errors and misconceptions that the learners may have, and tailor subsequent modeling activities to address these misconceptions in a timely manner. In fact, as a well-known YouTuber \citep{derekTEDed,derekPhD} proposes, and the 5E~instructional model supports \citep{BSCS5E}, addressing the students' \emph{misconceptions} about a topic is paramount in teaching science. In this regard, computational \emph{modeling} tools, as the vehicle for \gls{pbl} in engineering, enables the instructor to tailor the problems and questions posed to the students so that they can find, and explain, their own misconceptions.

For the reasons stated above, we have developed a new tool, and set the methodologies that will enable us, to conduct a pilot experience on implementing a computer-supported more active teaching of the subject of rocket engines to the students of the Space Systems MSc program taught at the \gls{upm}, in which the aim is to evaluate the impact ---both as the students' \emph{subjective} perception of the knowledge they acquire as well as by \emph{objectively} contrasting their grade of retention conducting tests to measure their performance--- of scientific inquiry-driven learning processes on the academic success of the learners.

The contribution of this work is, thus, on the one hand, a computer tool that addresses the lacks of currently available software in the field of rocket engines modeling and simulation specifically targeting the engineering classroom; and, on the other, guidelines for the development of such educational software. A computer tool, developed with almost the sole goal of facilitating the implementation of active learning strategies in teaching rocket engines to engineering students. And guidelines, provided as a list of `desirable features' ---that remain valid in principle irrelevantly of the field, not only within the subject of rocket engines--- that these tools should have, based on the available literature and upon our own developing and teaching experience, in order to be suitable to realize the full potential of active learning \emph{with} technology.

%%%%%%%%%%%%%%%%%%%%%%%%%%%%%%%%%%%%%%%%%%%%%%%%%%%%%%%%%%%%%%%%%%%%%%
%%%%%%%%%%%%%%%%%%%%%%%%%%%%%%%%%%%%%%%%%%%%%%%%%%%%%%%%%%%%%%%%%%%%%%
%%%%%%%%%%%%%%%%%%%%%%%%%%%%%%%%%%%%%%%%%%%%%%%%%%%%%%%%%%%%%%%%%%%%%%

\section{Tool Development}\label{sec:devel}

Prior to the beginning of the tool development itself, it is paramount to systematically establish concisely \emph{what} is it that we want to develop. From the literature review in Section~\ref{sec:motivation}, and a closer look into the readily available tools in the sector of rocket engines modeling, key features to have implemented in our tool as well as the general ``coding philosophy'' to abide by in the development process must be precisely defined.

\begin{figure}[t]
    \begin{center}
        \includegraphics[width=.45\linewidth,keepaspectratio=true]{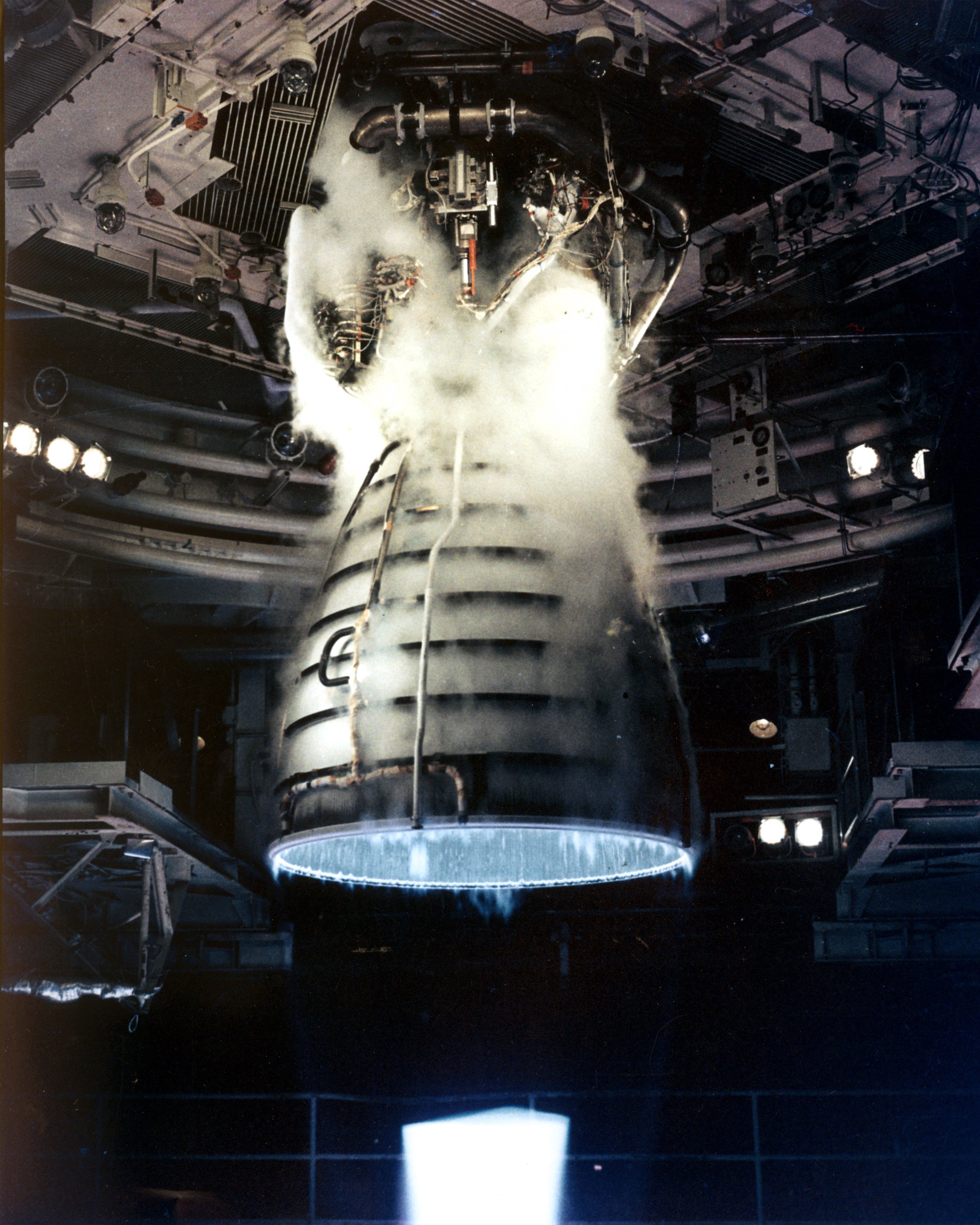}
        \caption{\gls{ssme} hot-firing test}\label{fig:ssme:firing-test}
    \end{center}
\end{figure}

\begin{figure}[t]
    \begin{center}
        \includegraphics[width=.95\linewidth,keepaspectratio=true]{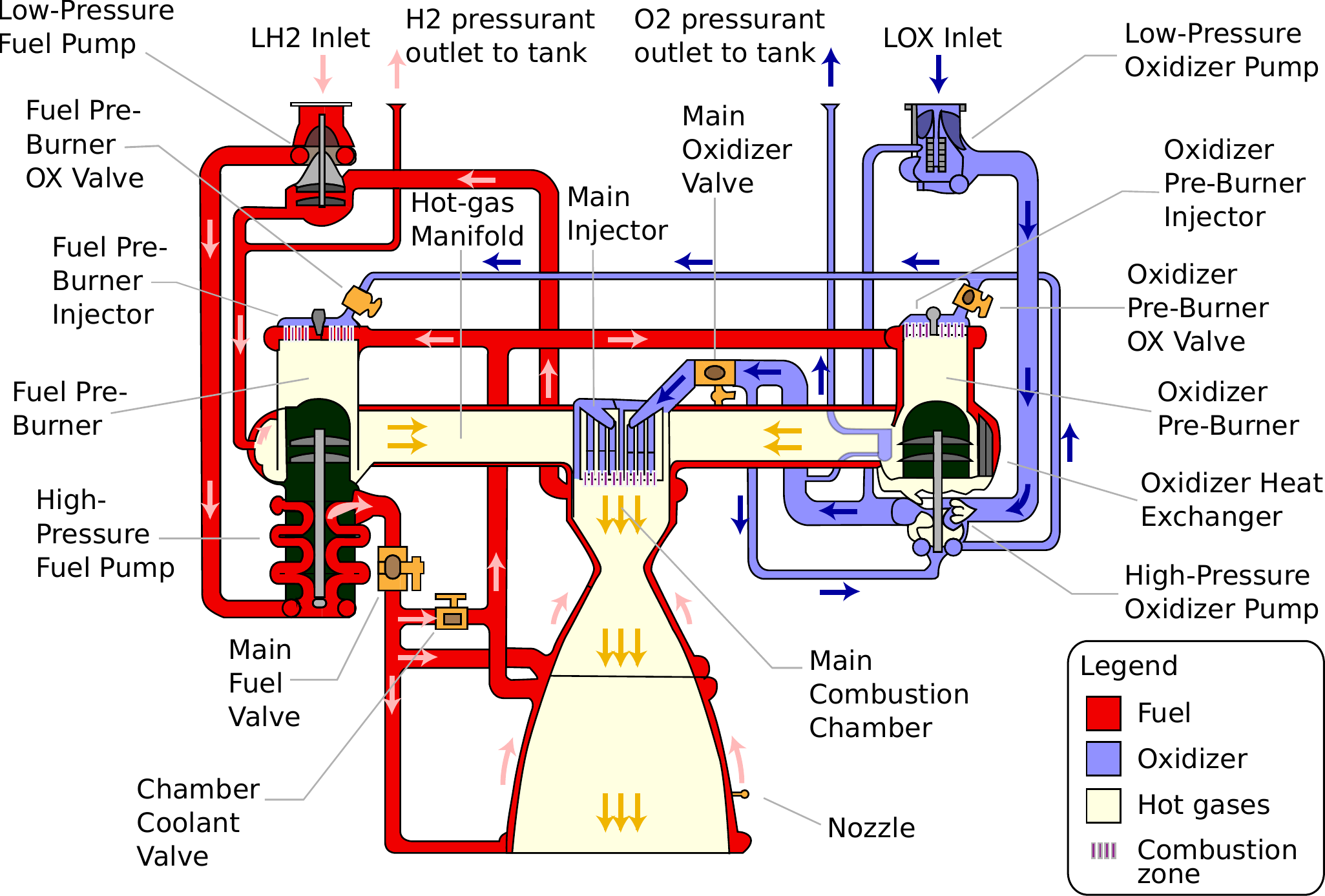}
        \caption{\gls{ssme} schematic diagram}\label{fig:ssme:diagram}
    \end{center}
\end{figure}

\Glspl{lprm} are efficient systems for aerospace propulsion and, as such, their study constitutes a substantial part of the academic curriculum of any MSc program taught within the aerospace sector. \Glspl{lprm} can be configured in many different ways, depending on the application; but, in any case, their performance is the result of the collective work of a significant number of specialized elements (pumps, turbines, burners, valves, etc.) organized by hydraulic networks arranged in a complex architecture with a large number of forks and junctions. Moreover, the fluids have properties that depend on the temperature conditions, they may undergo phase changes and be composed of mixtures originating in combustion processes. Indeed, the whole system is meant to feed the stored propellant into a combustion chamber and nozzle. An example of such a \gls{lprm} can be found in the \gls{ssme}, shown in figure~\ref{fig:ssme:firing-test}. The \gls{ssme} delivers up to 2.3~million Newton in thrust, and was used to propel the US Space Shuttle. The propellants were cryogenic liquid hydrogen and liquid oxygen, following a complex cycle (see figure~\ref{fig:ssme:diagram}) with a significant number of elements: five turbopumps, driven by one liquid turbine plus three gas turbines (two of which run on hot gases from the pre-burners, and hence need to be cooled), two heat exchangers, three combustion chambers, five hydraulically-actuated valves to control engine's output, etc. Thus it is made clear the enormous complexity of the system under study and the challenges and difficulties that trying to analyze the performance of such engine ``by hand'' would pose.

%\begin{figure}[t]
%    \begin{center}
%        %%%\hspace*{\fill}%%%\hfill
%        \begin{subfigure}[b]{.33\linewidth}
%            \begin{center}
%                \includegraphics[width=\linewidth,keepaspectratio=true]{SSME_firingTest}
%                \caption{}\label{fig:ssme:firing-test}
%            \end{center}
%        \end{subfigure}
%        \hfill%%%\hspace*{\fill}
%        \begin{subfigure}[b]{.66\linewidth}
%            \begin{center}
%                \includegraphics[width=\linewidth,keepaspectratio=true]{SSME_schematic}
%                \caption{}\label{fig:ssme:diagram}
%            \end{center}
%        \end{subfigure}
%        %%%\hspace*{\fill}%%%\hfill   % al final de la línea, "\hfill" no funciona bien...
%        \caption{\gls{ssme} hot-firing test~(\subref{fig:ssme:firing-test}) and diagram~(\subref{fig:ssme:diagram})}\label{fig:ssme}
%    \end{center}
%\end{figure}

From the teaching point of view, the practical activities carried out in the traditional classroom related to these problems often consist of simple exercises that can be solved with ``pen and paper''. The intention is to apply the knowledge and skills acquired by the learners to solve a real case, but the simplicity of the tools undermines the purported goal, and the student, because of the contrast with the reality facing him, may have the perception that what they are doing is trivial, oversimplified and non-worthwhile for their training. The present project aims to address this issue by making use of computer tools specifically tailored for the classroom environment ---i.e. with a flat learning curve, easy to grasp its fundamentals in a short period of time, and implementing models and concepts similar to the ones the student would use in solving the classroom problems ``by hand''. There are two types of benefits to be had with this approach: one is to improve the perception that the student has on what they do in the classroom, bringing their activity closer in line to their idea of what they will do in the real working world, which creates a positive predisposition to assimilate the concepts that these simple models can help to make clearer; the other is to allow the learner to take on more complex problems that could not be done otherwise ``by hand'', avoiding oversimplifications, which helps them to better judge and understand the behavior and performances of the systems simulated.

Implementing any sort of computer-supported active learning strategy in teaching \glspl{lprm}, as discussed in Section~\ref{sec:motivation}, means overcoming a few challenges, the first of which is to find the right software tool for \emph{modeling} (and not merely \emph{simulating}) these systems. As mentioned, most of the industry use their own in-house tools, inaccessible for the students, and specifically tailored for that particular company's very own products, but unable to tackle `generic modeling' problems in a flexible way. A few instances of different tools created for the simulation of \glspl{lprm} have been reported in the open literature, but these are mostly \emph{ad hoc} models, created specifically for the study of a particular effect in a very particular engine configuration. Among the initiatives to create flexible tools for the modeling of \glspl{lprm}, the LiRA project \citep{liraProject} and REDTOP-2 \citep{redtop2Project} stand out. However, in spite of the efforts made, it has to be remarked that both these tools are intended for \emph{simulation} of some configurations only, rather than allowing for actual \emph{modeling} of complex systems: both tools include a relatively extensive set of pre-defined configuration options through which different \gls{lprm} power plants can be analyzed, but constructing your own system, outside of those pre-defined configurations, is not allowed \citep[a more in-depth literature review on the different tools available can be found in][]{sierraLPRES}.

The few generic-purpose modeling tools commercially available, among which probably Simulink\supers{\textregistered} and EcosimPro\supers{\textregistered} are the most well-known, are often times used alongside their own in-house codes by companies in the aerospace sector and are, therefore, an option to consider. From the two, Simulink\supers{\textregistered} is more of a model-anything tool, whilst EcosimPro\supers{\textregistered} is the tool specifically tailored for the modeling and simulation of \glspl{lprm} by means of the \gls{espss} libraries ---particularly, the \gls{esa} has chosen EcosimPro\supers{\textregistered} as its recommended tool for simulation in several fields, including space propulsion, \glspl{eclss} and energy systems \citep[see, for instance,][]{espssRefManual,espss2008}--- and hence is the most widely used within the industry, so that it would seem more natural to teach the students of the Space Systems MSc program the tools that are specific to their sector, namely EcosimPro\supers{\textregistered}. Nevertheless, the \gls{espss} libraries for EcosimPro\supers{\textregistered} are a very complex set of models, which require specialized training to be able to use them, and have a steep learning curve that makes them not suitable at all for a classroom environment.

For these reasons, we have decided to develop a new tool, called \gls{lpres}, a simplified library for EcosimPro\supers{\textregistered} presented in \citep{sierraLPRES}. The tool is, therefore, a simplified version of the \gls{espss} professional libraries, developed mainly for educational purposes. The approach of implementing a `simplified' version of a more complex software tool for the classroom is not new: there are examples of similar strategies in other fields where accurately modeling the phenomena involved can result in highly demanding software tools \citep{woolsey2015}. But in our case, it has the additional benefit of providing a platform to get the students of the Space Systems MSc program to familiarize themselves with an environment ---i.e., EcosimPro\supers{\textregistered}--- widely used in companies within the aerospace sector, which will likely prove to be a valuable \emph{technological literacy} \citep{robotProof} for the learners to acquire during their training, at the same time that gives them the feeling of being `connected' to current standard practices in the industry, so that they perceive what they are being taught as useful and meaningful for their future professional careers. Furthermore, the intuitive ``drag and drop'' interface used by EcosimPro\supers{\textregistered}, along with its powerful symbolic and numeric solver, removes the burden from the student of having to deal with complex input configuration files and/or programming sophisticated methods for reaching numerical convergence of said models, so that the emphasis can be placed upon the \emph{modeling} task itself; much in line with the idea of creating a software tool that alleviates the `cognitive burden' from the students and lets them focus on the data gathering and interpretation processes \citep{campbell2010}.

Moreover, developing such a tool as \gls{lpres} with an educational mindset has other benefits to highlight. For starters, making the code available for the students (rather than having a palette of blocks functioning as `black-boxes' to choose from when building up the model) enables the learners to actually understand the relations and phenomena that govern the performance of the \gls{lprm}. The capability to choose from several degrees of pre-implemented model fidelity, or even to modify the code by themselves to account for new effects that they may consider relevant, allows the students to understand the \emph{limitations} and \emph{applicability} of the equations and models that they are dealing with, a fundamental goal of teaching \emph{modeling} according to \citep{schwarz2009,schwarz2005}. At the same time, it provides a meaningful platform in which to compare the results of the more simplified equations against the greater accuracy obtained by means of more complex modeling, showcasing the great value of being able to make quick \emph{estimates} about a particular system, a valuable ability for the engineer to have from the design point of view according to Dym~et~al. \citep{dym2005}. Furthermore, the possibility to re-arrange the hydraulic network of components of a particular \gls{lprm}, or to even create their very own system from scratch exploring novel configurations, provides the learners with the \emph{systems thinking} holistic view and the understanding of \emph{system dynamics} of the \glspl{lprm} that an engineer must have \citep{dym2005,robotProof}. Last, but not least, the way the EcosimPro\supers{\textregistered} modeling environment works, by means of setting up (\emph{virtual}) experiments, teaches the students how to \emph{design experiments} \citep{dym2005} so that they provide meaningful data, and teaches them how to interpret the results and infer conclusions from gathered data, enhancing their \emph{data literacy} \citep{robotProof}.   % Engagement from the students in long-term projects and willing to `know more'. Increased feeling of `ownership' and encourages learners to get actively involved (many came asking about available projects to be able to contribute to the development of the tool). A lo mejor mencionarlo aquí???

A relevant note on the benefits of developing our own tool must also be made herein. One of the shortfalls of the \gls{jitt} methodology is that requires extremely competent instructors, who are experts on the topics covered so that they can tailor the contents taught to the needs of the classroom \emph{on the fly} \citep{prince2006}. Much the same applies when the \gls{jitt} is made use of within the context of computational modeling, as in \citep{rodriguez2013}. In this regard, having our own tool ensures that we are experts on its usage and know its inner functioning in and out, which has allowed us to be effective in tailoring the modeling activities to address learners' misconceptions by guiding the inquiry-driven learning process. Making use of other available tools for which the instructors do not possess this kind of deep knowledge poses a risk to the effectiveness of such concepts as the \gls{jitt} and \gls{pbl}, since the instructor may not be able to know if (or how) a certain model to address a specific topic could be carried out with such a tool.

Lastly, it is worthwhile mentioning that developing \gls{lpres} not as a stand-alone tool, but rather as a library of components for EcosimPro\supers{\textregistered}, and putting the development of the code in hands of the students rather than being done by the professors, has facilitated over the past couple of years the \emph{cooperative education} concept by Aoun \citep{robotProof}, in which the university and the industry both cooperate and are responsible for the student's training. Empresarios Agrupados, the company in charge of developing EcosimPro\supers{\textregistered} embraced this project as theirs, and provided the opportunity for the students working on the \gls{lpres} library to be immersed in the professional world, by means of semester-long remunerated internships, so that the students could develop the library within an actual professional team of developers and had the chance to actually see, and get familiar with, the working environment by themselves, which will hopefully help them in making the best decisions and drive their future careers towards their goals with a more educated view on the topic.   %%%% Empresarios Agrupados, the company in charge of...   % for the double-blind review, HIDE THIS !!!

%%%%%%%%%%%%%%%%%%%%%%%%%%%%%%%%%%%%%%%%%%%%%%%%%%%%%%%%%%%%%%%%%%%%%%

%\section{The EcosimPro\supers{\textregistered} Computer Program}\label{sec:EcosimPro}
% alternatively:
%   what comes within the \texorpdfstring{} goes as the title itself, where you can call \Gls, \hl, etc.
%   what comes afterwards between {} goes into the table of contents, index, etc. and must be 'plain' text:
%                                                               for that matter, \textregistered would be allowed as 'plain' text, just not within a \supers{}
\subsection{\texorpdfstring{The EcosimPro\supers{\textregistered} Computer Program}{The EcosimPro Computer Program}}\label{sec:EcosimPro}

EcosimPro\supers{\textregistered} is a powerful simulation environment capable of modeling any type of dynamic system represented by \glspl{dae}, \glspl{ode} and/or discrete events. It is based on numerical and symbolic methods that can process complex systems of algebraic and differential equations. It is endowed with an intelligent system for problem analysis that lets the user pose tough problems and adopt optimal solution strategies. EcosimPro\supers{\textregistered} has a very complete and intuitive graphical user interface where all the work is carried out, from building the model by dragging and dropping component icons from several palettes within pre-implemented component libraries into a schematic window, to programming the behavior of any element by means of its very own non-causal language, called \gls{el}.

EcosimPro\supers{\textregistered} is used in companies in the energy and aerospace sectors, and particularly the \gls{esa} has chosen EcosimPro\supers{\textregistered} as its recommended tool for simulation in several fields \citep[see][]{espssRefManual}, including space propulsion \citep[see][]{espss2008}, \glspl{eclss} and energy systems.   %%%% In the aerospace sector, there is the \gls{proosis} program. It is based on EcosimPro\supers{\textregistered}, taking all its capabilities along with additional ones geared to simulating gas turbines and other aeronautical systems. \Gls{proosis} includes the TURBO library \citep[see][]{proosis2007}, with complex models for gas turbines and other components such as compressors, shafts and heat exchangers, and is being used by major international firms in the sector.

%%%%%%%%%%%%%%%%%%%%%%%%%%%%%%%%%%%%%%%%%%%%%%%%%%%%%%%%%%%%%%%%%%%%%%

%\section{\glsentryshort{espss} Professional Libraries}\label{sec:ESPSS}
% alternatively:
%   what comes within the \texorpdfstring{} goes as the title itself, where you can call \Gls, \hl, etc.
%   what comes afterwards between {} goes into the table of contents, index, etc. and must be 'plain' text
\subsection{\texorpdfstring{\Gls{espss} Professional Libraries}{\glsentryshort{espss} Professional Libraries}}\label{sec:ESPSS}

The purpose of the \gls{espss} libraries is to provide a standard set of components and functions to simulate jet propulsion systems and space propulsion systems. The \gls{espss} also provides a standard database for propellants, pressurization substances and other fluids. The \gls{espss} libraries solve the continuity, momentum and energy equations in a discretized mesh that can vary depending on the component, but that in general terms can be understood most of the times as a 1D-CFD. It is capable of handling multi-phase flow, it retains all the transient terms in the equations, and certainly the working fluids are treated as real gases/liquids, with properties depending both on the temperature and pressure conditions. Needless to say, the associated cost with this fine level of simulation detail is the vast amount of information that the user must provide as an input: the full geometry of the whole system must be known beforehand, so that volumes in the fluid network can fill up or empty out; construction materials for pipes or tanks must be specified, so that heat transfer calculations and elasticity/elongation of certain elements can be accounted for; etc. This sort of approach, useful in the industry, is not appropriate for a classroom environment, where the students might feel overwhelmed by the sheer amount of knowledge about a particular system that is required from the user in order to be able to build up a single model.   %%%% The \gls{espss} libraries belong to the \gls{esa} and their use is subject to prior approval from the \gls{esa}, with Empresarios Agrupados Internacional being the developer and official distributor for \hl{outside customers??}.

\begin{figure}[t]
    \begin{center}
        \includegraphics[keepaspectratio=true,width=.95\linewidth]{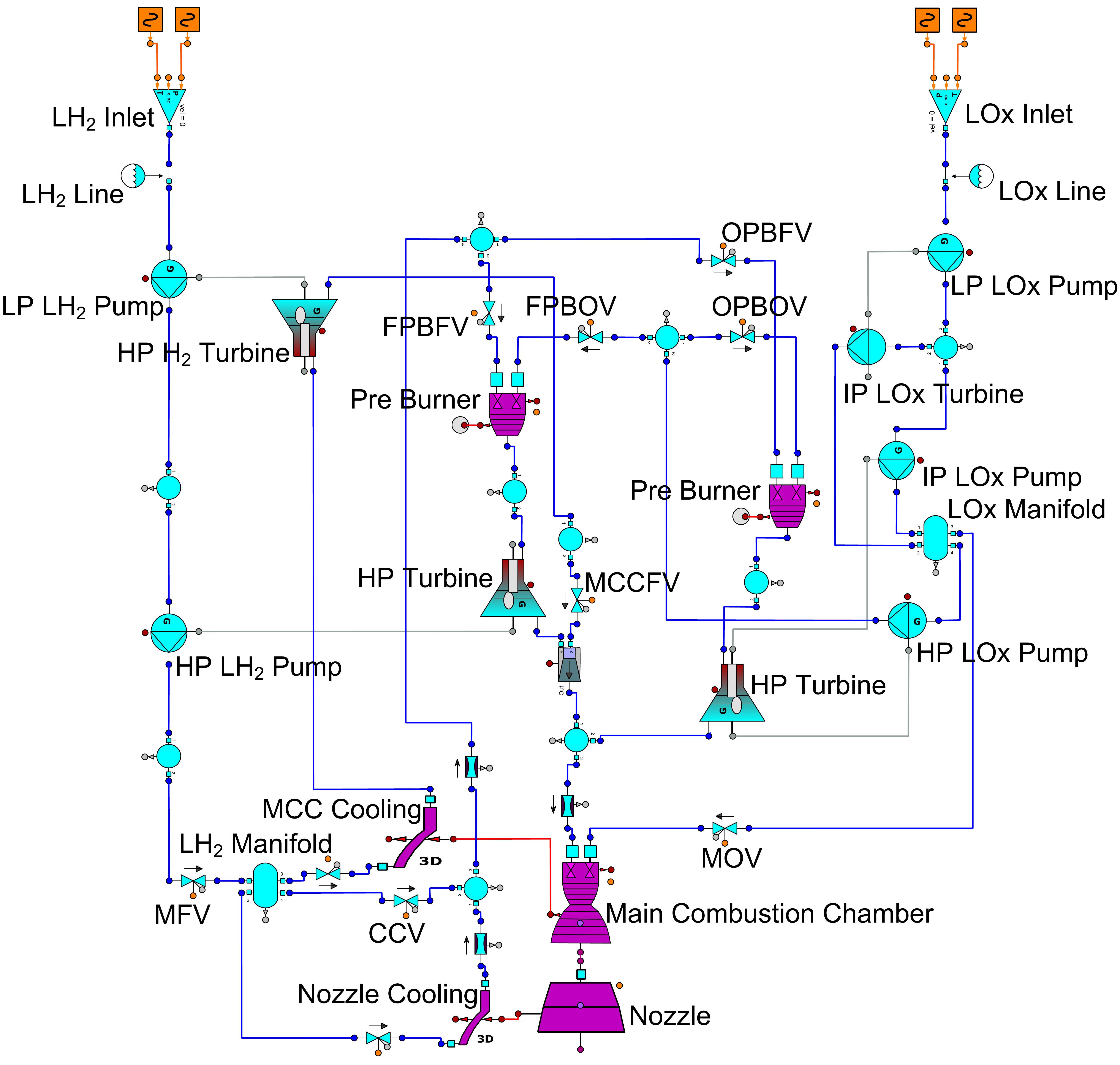}
        \caption{Example of \gls{lprm} modeled with the \gls{espss} libraries.}\label{fig:ssme-espss}
    \end{center}
\end{figure}

Figure~\ref{fig:ssme-espss} shows a typical schematic created with \gls{espss} for simulating an engine. In this case it is a closed-cycle turbo-powered bi-propellant rocket motor. More specifically, it is a staged-combustion cycle. The program can carry out direct calculations of the steady-state equations for an engine \citep[see][]{ariane2010}, perform transient analyses regarding the startup, thrust control and shutdown of the system \citep[see][]{espss2010}, and even be programmed to do optimization tasks \citep[see][]{espss2012}. This type of work is customary in designing and analyzing these complex engineering systems.

Companies usually have their own software for analyzing their products, and modify it depending on the changes they make to their components, the innovations they add to their calculation methods and the types of simulation they end up doing on a project. Some of the advantages of the \gls{espss} libraries over the in-house software specific to a product include the ease for developing new models (or improving existing ones), the flexibility for adapting the same tool to a large number of different configurations simply by using standard components from a palette, the evolution and improvement of these libraries thanks to projects carried out every year by the industry, and the ease for exchanging models between companies (or, for that matter, between centers within the same company) by having a common tool.

%%%%%%%%%%%%%%%%%%%%%%%%%%%%%%%%%%%%%%%%%%%%%%%%%%%%%%%%%%%%%%%%%%%%%%

%\section{\glsentryshort{lpres} Educational Library}\label{sec:LPRES}
% alternatively:
%   what comes within the \texorpdfstring{} goes as the title itself, where you can call \Gls, \hl, etc.
%   what comes afterwards between {} goes into the table of contents, index, etc. and must be 'plain' text
\subsection{\texorpdfstring{The \Gls{lpres} Educational Library}{The \glsentryshort{lpres} Educational Library}}\label{sec:LPRES}

\Gls{lpres} is conceived to imitate the capabilities of the professional \gls{espss} libraries. As the \gls{espss}, the \gls{lpres} library provides a set of standard re-usable components, with which the user can model virtually any imaginable jet propulsion or space propulsion system (ranging from aeroengines to turbo-charged \glspl{lprm}) by mere ``drag and drop'', or upon which the user may further expand its capabilities by implementing models for new components or refining the modeling of currently existing ones. The \gls{lpres} library has been developed to make use of the powerful solver of EcosimPro\supers{\textregistered}, which enables the learner to focus on the \emph{modeling} aspect, leaving the numerical hassle to be dealt with by the program.

It must be remarked herein that the intention is \emph{not} to code an exact replica of (nor any sort of replacement for) the \gls{espss} libraries, but rather something else. The main goal is the same: to model and simulate jet propulsion systems and space propulsion systems. But the means are substantially different, given that the \gls{lpres} library is conceived for the classroom environment, with the emphasis placed upon the educational usage of the tool, rather than on the accuracy of the numerical results. In this regard, one of the main objectives is to reduce as much as possible the amount of information that the user must input into the models, so that the modeling activity seems affordable for the students at the same time that makes up for a much smoother learning curve. The characterization of the components is thus reduced ---by means of simplified formulation based on first principles only, very similar to the formulation that one would use when solving problems with ``pen and paper'' (neglecting effects that can be considered of second order, and focusing only on the steady-state operation of the components, for instance)---, most of the times, to a mere characteristic length (or radius) and/or a free-flow cross-sectional area. The need for a discretized mesh is therefore eliminated, and the formulation is reduced from the 1D-CFD of \gls{espss} to just a fluid network of 0D components in \gls{lpres}. Furthermore, as the intention is just to obtain ``\emph{qualitatively} good results'', useful maybe for the very preliminary design phase of a \gls{lprm} cycle, but not 100\%{} accurate numerical values, the formulation of the components is simplified even more by considering ideal liquids and ideal gases only, with ideal combustion products, so that complex calculations on the chemical equilibrium of the substances involved are avoided altogether.

To make up for these strong simplifications, the components include a number (this can vary from every component to one another) of degrees of freedom, mostly in the form of \emph{efficiencies}, that allow the user to fine-tune a model to replicate the performance of any known system. Hence, the modeling process comprises two steps: firstly, the user \emph{aligns} the model with the actual performances of the system they want to replicate, in what is called the \emph{design} step; then, they can proceed forward to simulate that system at any \emph{off-design} operating condition that they may so desire. The \emph{design} step serves both to determine the \emph{efficiencies} that would make a particular model with the simplified formulation perform closely enough to the available real data, and to perform the basic \emph{sizing} of the geometry (this is, to calculate the characteristic lengths, radii and free-flow cross-sectional areas of the different components) so that \emph{off-design} calculations (for which both the efficiencies and the geometry are kept constant) can be tackled subsequently. More details about the formulation used within each component can be found in Sierra~et~al.~\citep{sierraLPRES}.

With this approach, the two main challenges stated in Section~\ref{sec:motivation} have been successfully addressed: on the one hand, the expected degree of realism for describing the different elements is achieved by means of the aforementioned efficiency parameters, that enable the simplified components to be fine-tuned to replicate the behavior of a real system; and, on the other, the complexity ---understood as the presence of many interrelated components--- of the systems to analyze is dealt with by the powerful symbolic--numerical solver of EcosimPro\supers{\textregistered}.

\begin{figure}[t]
    \begin{center}
        \includegraphics[keepaspectratio=true,width=.95\linewidth]{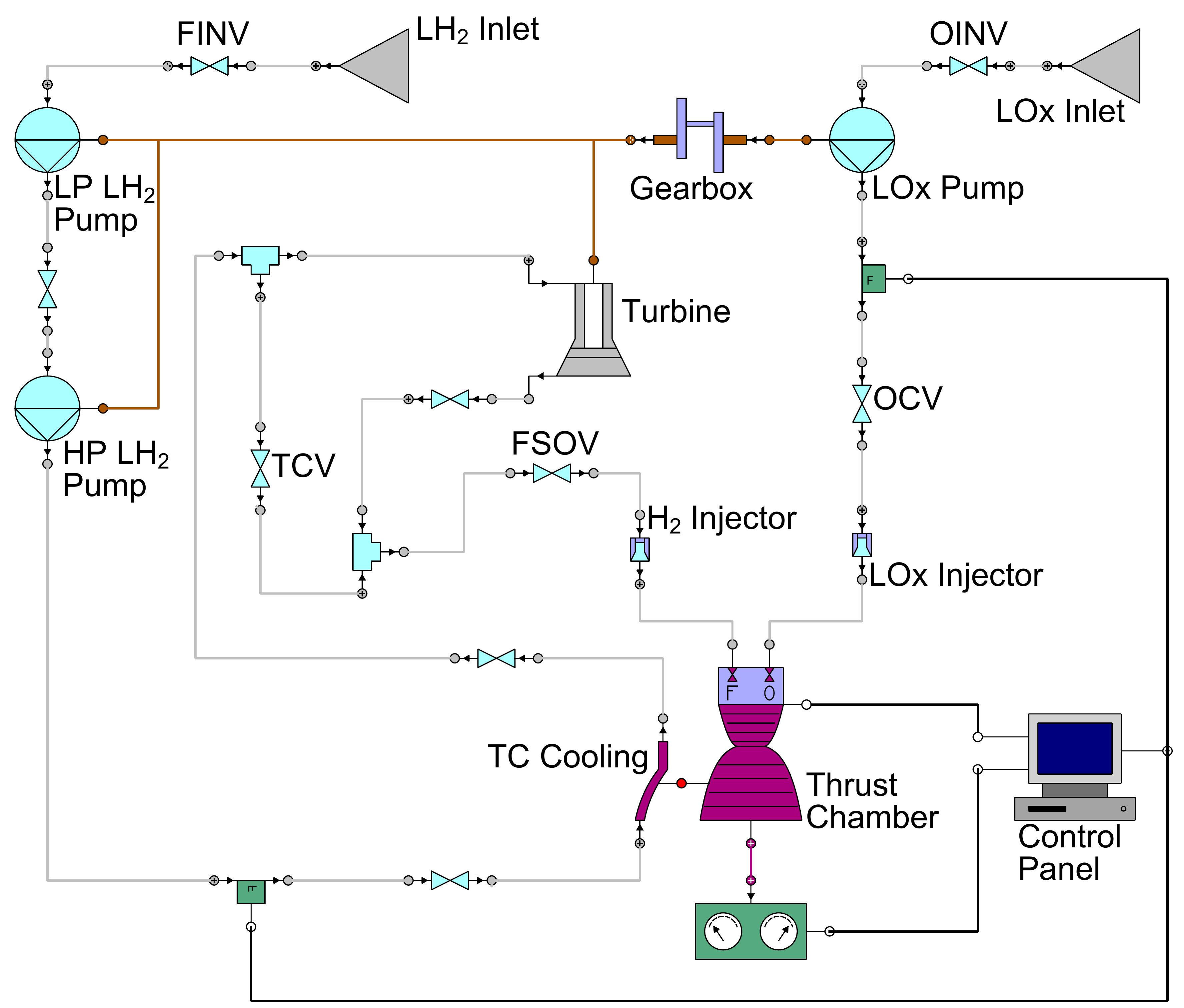}
        \caption{Example of a model with the \gls{lpres} library. Simulation of the \mbox{RL10A-3-3A} rocket engine \citep{rl10a33a}.}\label{fig:lpres_rl10-full}
    \end{center}
\end{figure}

Figure~\ref{fig:lpres_rl10-full} shows an example of modeling with \gls{lpres} in which a closed-cycle bi-propellant rocket engine has been modeled. Specifically, it is an expander cycle belonging to rocket engine \mbox{RL10A-3-3A} \citep{rl10a33a}. The propellants are stored in two tanks: one for the oxidant and the other for the fuel. Two pumps are in charge of separately pressurizing the propellants. Then, the fuel goes through some channels used for cooling the nozzle. While circulating through the cooling system, the fuel heats up, thereby increasing its enthalpy enough to move a turbine located at the outlet of the nozzle cooling system. This turbine is the one in charge of powering the pumps. After passing through the turbine, the fuel is injected into the combustion chamber along with the oxidant. Finally, there are several metering elements to calculate characteristics achieved by the whole assembly, such as the thrust and the specific impulse.

Figure~\ref{fig:lpres_rl10-full} illustrates how intuitive the modeling of complex systems becomes with \gls{lpres} for the learners, so that introducing modifications into the model, or building up new models from scratch, does not pose a problem for implementing \gls{pbl} strategies in the classroom. It also comes to show the similarities between the usage of the \gls{espss} professional libraries in figure~\ref{fig:ssme-espss} and the \gls{lpres} educational libraries in figure~\ref{fig:lpres_rl10-full}, both having a very similar appearance, so that the students do get the feeling that they are working with a professional tool, part of the standard practices within the industry today. The relevant documentation produced for the \gls{lpres} library does also follow current industry standards, much in the line of the \gls{espss} manuals, to reinforce this perception.   % removed the figure with the manuals...

%%%%%%%%%%%%%%%%%%%%%%%%%%%%%%%%%%%%%%%%%%%%%%%%%%%%%%%%%%%%%%%%%%%%%%

%\subsection{Components of the \glsentryshort{lpres} Library}\label{sec:LPRES:components}
% alternatively:
%   what comes within the \texorpdfstring{} goes as the title itself, where you can call \Gls, \hl, etc.
%   what comes afterwards between {} goes into the table of contents, index, etc. and must be 'plain' text
\subsubsection{\texorpdfstring{Components of the \gls{lpres} Library}{Components of the \glsentryshort{lpres} Library}}\label{sec:LPRES:components}

The \gls{lpres} library consists of several models representing the basic components making up a \gls{lprm}. However, unlike the professional \gls{espss} library, since the \gls{lpres} library uses simplified models taught in the classroom, it is steady-state only, which means all its components have null derivatives with respect of time in their equations. 

\begin{figure}[t]
    \begin{center}
        \includegraphics[keepaspectratio=true,width=.95\linewidth]{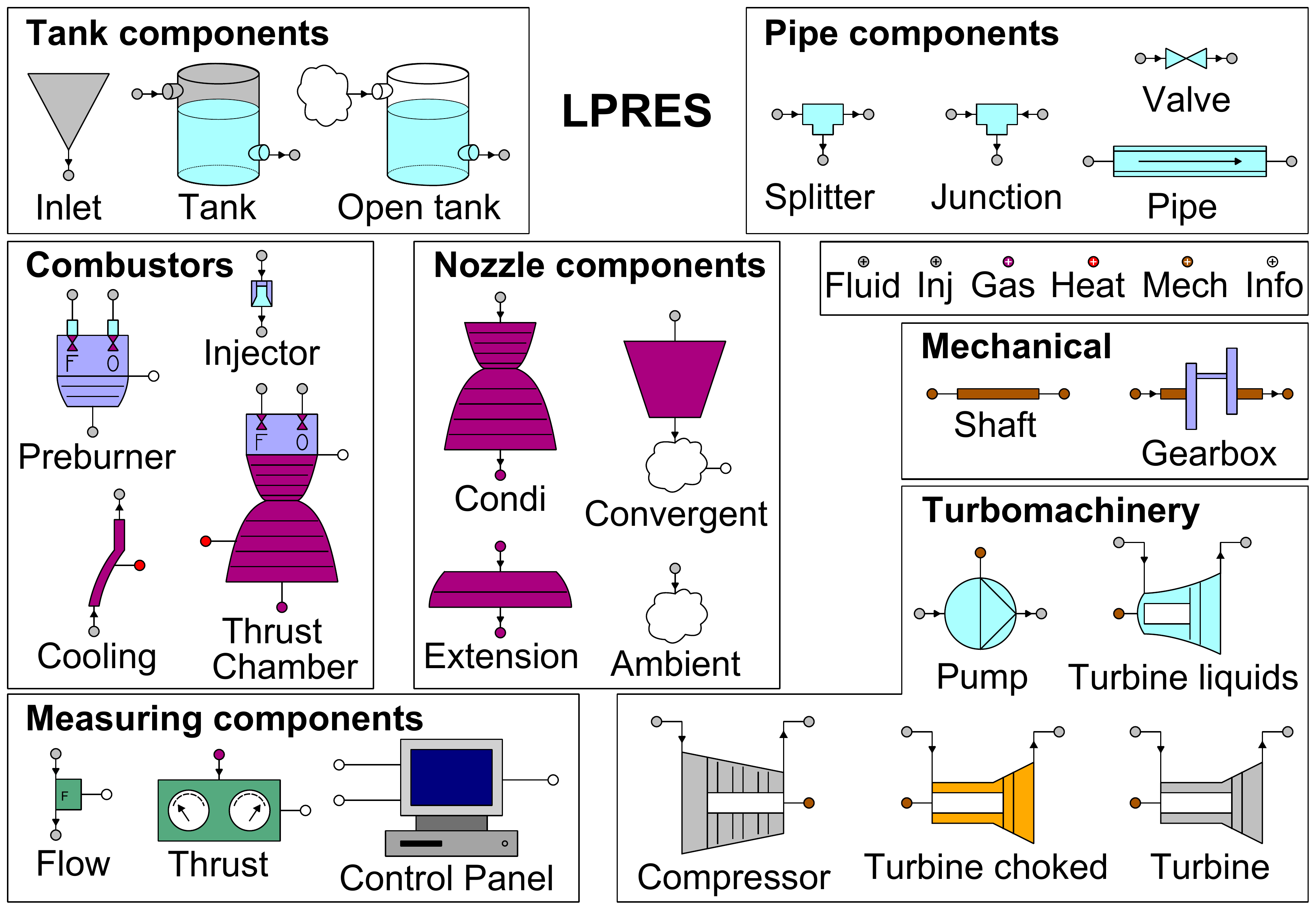}
        \caption{Available components within the \gls{lpres} educational library.}\label{fig:lpres_tree}
    \end{center}
\end{figure}

Figure~\ref{fig:lpres_tree} shows the components that the \gls{lpres} library makes available for the user. Ports are essential elements, since they are in charge of connecting any component to one another, by means of the right interface that specifies which variables are passed on. The library relies heavily on having the right definition for the connection ports to allow for versatile and efficient modeling. Different types of tanks and the components needed to perform the fluid dynamic connections between elements are made available, too. Components that represent turbomachinery as well as those from the mechanical union between turbines, compressors and pumps are included within this library. The remaining components shown represent combustion chambers, nozzles (convergent and convergent-divergent), nozzle cooling circuits and injectors, as well as the measuring components used to predict how the rocket engine will act as a whole. They can all be used to create models for more complex systems by means of simple ``drag and drop'' actions from the palette of re-usable components.

%%%%%%%%%%%%%%%%%%%%%%%%%%%%%%%%%%%%%%%%%%%%%%%%%%%%%%%%%%%%%%%%%%%%%%

%\subsection{Examples with the \glsentryshort{lpres} Library}\label{sec:LPRES:examples}
% alternatively:
%   what comes within the \texorpdfstring{} goes as the title itself, where you can call \Gls, \hl, etc.
%   what comes afterwards between {} goes into the table of contents, index, etc. and must be 'plain' text
\subsubsection{\texorpdfstring{Examples with the \gls{lpres} Library}{Examples with the \glsentryshort{lpres} Library}}\label{sec:LPRES:examples}

Several examples were built with \gls{lpres} in EcosimPro\supers{\textregistered}, mainly to test the robustness of the library, and are now bundled within the same package. Making sure that the code is robust enough (so that modeling novel configurations, and getting meaningful results from these, without having issues with the numerical aspects of solving complex systems of equations) is paramount in facilitating the learners' autonomous usage of the tool. However, this is not their only purpose, since these examples also help to ease the user's learning process by a great deal, given that they are all readily available for the students and are fully working and error-free models that can be used to showcase the capabilities of the library.

There are two types of models within the package: what we have called the (simpler) ``test cases'' and (more complex) ``whole-engine examples'', all of them inspired (or closely resembling) actual ``real-life'' rocket engines. We take from Roache \citep{roacheValVerif} the succinct definition of \emph{verification} as ``solving the equations right'' and \emph{validation} as ``solving the right equations''. Simple test cases (modeling, for the most part, sub-systems of actual \gls{lprm} cycles) were used to verify the components. Furthermore, as these test cases contain just a few components within each model, they can also prove highly useful to the learner in understanding the modeling equations and behavior of some of these components when they are isolated from everything else (or just connected to very few other elements in simple fluid networks). Whole-engine examples, on the other hand, feature much more complex network arrangements and have many interrelated components within each model, and hence were used to validate the library. These whole-engine examples comprise some of the most representative cycles that can be modeled with the \gls{lpres} library: a gas generator cycle, an expander cycle (such as the one shown in figure~\ref{fig:lpres_rl10-full}), a pressurized rocket engine, and even a jet engine, are all included within. Moreover, some of these whole-engine examples were used to validate the library directly, against real publicly-available data, or indirectly, against results from other computational tools that had been previously validated themselves (such as the \gls{espss}). As a matter of fact, the expander cycle shown in figure~\ref{fig:lpres_rl10-full} was one of such whole-engine models used for validation (in this case, indirect validation against data from simulations with the \gls{espss} professional libraries). The most representative results from this validation case, which also serves to illustrate how the \gls{espss} professional libraries and the \gls{lpres} educational tool compare with each other, are summarized in table~\ref{tab:validation}.

\begin{table}[!t]
    \begin{center}
        \caption{Validation data for the design model of the \mbox{RL10A-3-3A} rocket engine, comparing the results obtained by means of the \gls{lpres} simplified formulation with the more accurate simulations performed by \gls{espss}.}
        \label{tab:validation}
        \begin{tabularx}{.95\linewidth}{p{.55\linewidth} *{3}{Y}}
            \toprule[1.5pt]\addlinespace[1.5pt]\midrule
              &  \Gls{lpres}  &  \Gls{espss}  &  rel. error  \\ \cmidrule(lr){2-2} \cmidrule(lr){3-3} \cmidrule(lr){4-4}
            Fuel injector discharge area  [cm\supers{2}]             &   17.8  &   17.7  &  \mbox{\phantom{0}0.6\%{}}  \\
            Oxidizer injector discharge area  [cm\supers{2}]         &    5.8  &    6.1  &  \mbox{\phantom{0}4.9\%{}}  \\
%%%            Convergent-divergent nozzle throat area  [cm\supers{2}]  &  134    &  116    &                  15.5\%{}   \\
%%%            Turbine inlet area  [cm\supers{2}]                       &    6.9  &    4.2  &                  64.3\%{}   \\
            Turbine mass flow rate  [kg/s]                           &    2.1  &    2.2  &  \mbox{\phantom{0}4.5\%{}}  \\
            Combustion chamber temperature  [K]                      & 3181    & 3242    &  \mbox{\phantom{0}1.9\%{}}  \\
            Cooling jacket outlet total temperature  [K]             &  164    &  174    &  \mbox{\phantom{0}5.7\%{}}  \\
            Cooling jacket outlet total pressure  [bar]              &   69.6  &   76.1  &  \mbox{\phantom{0}8.5\%{}}  \\
            \midrule\addlinespace[1.5pt]\bottomrule[1.5pt]
        \end{tabularx}
    \end{center}
\end{table}

The data shown in table~\ref{tab:validation} correspond to the \emph{design} calculations for the \mbox{RL10A-3-3A} rocket engine \citep{rl10a33a}, and compares the results obtained with the \gls{lpres} simplified models, against data from simulations with \gls{espss}. In \gls{espss}, the full geometry of the rocket engine has been provided as an input, as well as the boundary conditions, and the performances of the rocket engine have been calculated. In \gls{lpres}, on the contrary, some of the operational data of the system that we want to replicate are supplied as input data by the user (namely: oxidizer and fuel mass flow rates, pumps pressure leaps, turbine expansion ratio and rotational speed, etc.), and these data are used to \emph{align} the simplified model with the original system by means of adjusting certain efficiency parameters and calculating the characteristic geometry (characteristic lengths and/or radii, and free-flow cross-sectional areas) for the main components. Table~\ref{tab:validation} shows that the objective of getting qualitatively good results, in decent agreement with reality, has been achieved. Relative errors for the vast majority of calculated parameters is kept below 5\%{}, and only certain specific elements ---as might be the cooling jacket, that involves more complicated heat-transfer calculations--- inherently hard to model by means of 0D components, yield errors slightly higher (see, for instance, the temperature and pressure conditions at the outlet of the cooling jacket reported in table~\ref{tab:validation}, which both accumulate errors nevertheless well below 10\%{}). This sort of discrepancies are usually considered acceptable for the very preliminary design stages (configuration proposals) of any rocket engine, and are definitely considered sufficient for an educational tool aimed at developing the \emph{systems thinking} and \emph{design thinking} competencies of the students. %%% The relative errors that these calculations yield, when compared against the actual geometry supplied to \gls{espss} is below 5\%{} for the vast majority of the output geometric parameters; however, there are a few exceptions (most notably, the turbine inlet area, as the data in table~\ref{tab:validation} show) that incur in relative errors that can exceed even 60\%{} some times. It is important to keep in mind, though, that the geometry calculated by \gls{lpres} is not `actual' geometry, but rather `characteristic' geometry: that is to say, \gls{lpres} does not calculate the actual size of a turbine, but instead the characteristic size that a turbine that would behave as the simplified formulation predicts should have in order to replicate the actual operating conditions of the real system; and these two values are seldom comparable, especially for the turbomachinery components. The operational parameters, such as the temperature and pressure conditions shown in table~\ref{tab:validation}, nevertheless, should yield lower discrepancies. With the sole exception of the cooling jacket, that involves more complicated heat-transfer calculations, hard to model by means of 0D components, the rest of operational parameters yield errors below 5\%{} (and even the cooling jacket outlet conditions are calculated with relative errors well below 10\%{}).

It must be noted, though, that these relative errors ---i.e. in the calculated operational parameters--- might increase slightly when performing off-design simulations (generally speaking, the further the operating condition is from the set design point, the greater the errors that tend to accumulate), but are in any case kept below reasonable margins (typically well below 10\%{}) and, more importantly, the trends of the \emph{system dynamics} are always captured accurately from a qualitative point of view, which makes the tool again very useful to help the students develop their \emph{systems thinking} competencies.

%%%%%%%%%%%%%%%%%%%%%%%%%%%%%%%%%%%%%%%%%%%%%%%%%%%%%%%%%%%%%%%%%%%%%%
%%%%%%%%%%%%%%%%%%%%%%%%%%%%%%%%%%%%%%%%%%%%%%%%%%%%%%%%%%%%%%%%%%%%%%
%%%%%%%%%%%%%%%%%%%%%%%%%%%%%%%%%%%%%%%%%%%%%%%%%%%%%%%%%%%%%%%%%%%%%%

\section{Pilot Experience}\label{sec:pilot}

%Finding the right balance between the amount of time spent in the classroom and the usage of software tools that encourages and reinforces learning is not an easy task.

Once a sufficiently stable version of the \gls{lpres} library was developed, the next step was to implement a pilot experience in the classroom and to evaluate the initial results. The workshop carried out within the context of the MSc in Space Systems taught at the \gls{upm} consists of two different parts. First, the students are asked to attend a total of 8~hours of ``practical lectures'' in the computer lab; then, the students are given a short assignment, with an estimated overall dedication of an additional 8~hours on their own. Overall, the whole pilot experience, including both parts, spans over a total duration of 6 to 8~weeks.

The 8~hours of computer-lab ``practical lectures'' are used to gradually build up from a model of \emph{structured inquiry} \citep{prince2006} at the beginning, in which the instructor outlines not only the problem but also some guidelines for solving the questions posed, so that the learners get familiar with the tool first; to a model of \emph{guided inquiry} \citep{prince2006} towards the end, in which the students no longer need guidelines for the solution of the problem, since they are already confident in the usage of the software, and they can take on answering the questions posed by the professor on their own. The assignment follows the concept of \gls{pbl} \citep{prince2006}, formulating an open-ended ``real-world'' problem that requires the students not only to investigate a certain phenomenon or system, but also to make some `design decisions' and to learn how to deal with, and reason about, the \emph{uncertainty} associated with those decisions along the way \citep{dym2005}.

The set of major abilities that the students are expected to learn from this workshop is summarized in table~\ref{tab:abilities}. From these abilities, and as previously described, the competencies that the learners will be able to develop thanks to this sort of academic training are gathered in table~\ref{tab:competencies}.

\begin{table}[!t]
\begin{center}
\caption{Set of abilities learned by the students through the \emph{EcosimPro\supers\textregistered~|~\gls{lpres}} workshop.}
\label{tab:abilities}
\begin{tabularx}{.85\linewidth}{*{1}{X}  p{.75\linewidth}}
\toprule[1.5pt]\addlinespace[1.5pt]\midrule
1  &  Advanced usage of the EcosimPro\supers\textregistered environment and capabilities.                               \\ \midrule[0pt]    %%% --->  Q5
2  &  Development of new models and components by means of coding in \gls{el} within EcosimPro\supers\textregistered.   \\ \midrule[0pt]    %%% ---> Q10 & Q12
3  &  Modeling and simulation of complex \gls{lprm} cycles by using \gls{lpres} re-usable components.                   \\ \midrule[0pt]    %%% ---> Q11
4  &  Programming and running experiments to gather relevant data in EcosimPro\supers\textregistered~|~\gls{lpres}.     \\ \midrule[0pt]    %%% ---> Q14
5  &  Deep understanding of the operation and control of \glspl{lprm} and their essential sub-systems.                  \\                  %%% --->  Q3 & Q8
\midrule\addlinespace[1.5pt]\bottomrule[1.5pt]
\end{tabularx}
\end{center}
\end{table}

\begin{table}[!t]
\begin{center}
\caption{Main competencies developed by the learners through the \emph{EcosimPro\supers\textregistered~|~\gls{lpres}} workshop.}
\label{tab:competencies}
\begin{tabularx}{.85\linewidth}{*{1}{X}  p{.75\linewidth}}
\toprule[1.5pt]\addlinespace[1.5pt]\midrule
1  &  \emph{Systems thinking} \citep[as defined by][]{robotProof}, both in terms of \emph{systems design} as well as \emph{system dynamics} \citep[as per][]{dym2005}.      \\ \midrule[0pt]
2  &  \emph{Experiments design} \citep[as per][]{dym2005} \&{} \emph{data literacy} \citep[as per][]{robotProof}.                                                           \\ \midrule[0pt]
3  &  \emph{Technological literacy} \citep[as defined by][]{robotProof}.                                                                                                    \\
\midrule\addlinespace[1.5pt]\bottomrule[1.5pt]
\end{tabularx}
\end{center}
\end{table}

%%%%%%%%%%%%%%%%%%%%%%%%%%%%%%%%%%%%%%%%%%%%%%%%%%%%%%%%%%%%%%%%%%%%%%
%%%%%%%%%%%%%%%%%%%%%%%%%%%%%%%%%%%%%%%%%%%%%%%%%%%%%%%%%%%%%%%%%%%%%%
%%%%%%%%%%%%%%%%%%%%%%%%%%%%%%%%%%%%%%%%%%%%%%%%%%%%%%%%%%%%%%%%%%%%%%

\section{Results and Discussion}\label{sec:results}

Properly speaking, the evaluation of the impact on learners' performance of the usage of the developed tool in teaching the subject of rocket engines should be able to show how much better the learning objectives are achieved by comparing at least two groups: one in which this pilot experience is put into practice (the `experiment' group) and the other in which it is not (the `control' group). Due to the small number of students enrolled in the Space Systems MSc every year, though, strictly limited to 20, it was impossible to ensure that the two groups would not be biased, and thus the decision was made to try the `experiment' condition with the whole class.

As a means to objectively evaluate the impact of the usage of our developed tool in the classroom, a pre-workshop test was run, asking the students to answer a set of questions dealing with the topics that had already been covered during the theoretical lectures. This pre-test had a two-fold aim:
\begin{inparaenum}[\itshape a\upshape)]
\item to serve as a starting point to measure the retention from the students by means of only attending theoretical lectures in a mostly `passive' role, and the gains achieved by means of the practical computer modeling activities imparted thereafter; and
\item to help the instructor identify common error or misconceptions within the class, and tailor the problems posed to be investigated specifically to tackle those misconceptions, apart of the more general learning objectives established for the course.
\end{inparaenum}
A post-test was also carried out, essentially asking the students to answer very similar questions about the same topics, following already established methodology \citep[see, for instance][]{carey2017,YanLiYinNie2018}, to objectively measure the gains in retention.

Results for both the pre-test and the post-test are summarized in table~\ref{tab:tests}. As shown, the average score achieved by the students more than doubled from the pre-test to the post-test. Even though the samples are small, the Welch's unequal variances \textit{t}-test yields a \textit{p}-value~$<0.1\thinspace\%{}$ with a significance level $\alpha=0.1\thinspace\%{}$ (power of the statistic test using the assumption of equal variances for both samples close to~99\%{}), which showcases that the null hypothesis is extremely likely to be rejected and the gains in retention achieved by the students from the practical workshop can indeed be treated as significant for the purposes of this study, even for such a low value of $\alpha$.

\begin{table}[!t]
\begin{center}
\caption{Summary of results from the pre- and the post-test for the `experiment' condition ($N=17$; marks on a /100 scale).}
\label{tab:tests}
\begin{tabularx}{.4\linewidth}{p{0.15\linewidth} *{2}{Y}}
\toprule[1.5pt]\addlinespace[1.5pt]\midrule
             &   mean    &   st. dev.    \\ \cmidrule(lr){2-2} \cmidrule(lr){3-3}
pre-test     &   28.83   &   12.64       \\
post-test    &   61.51   &   15.95       \\
\midrule\addlinespace[1.5pt]\bottomrule[1.5pt]
\end{tabularx}
\end{center}
\end{table}

In addition to the above, a short survey was conducted among the participants after the workshop (detailed results from this survey can be found in table~\ref{tab:survey}), to evaluate their subjective perception of the overall impact of this training on their whole academic program. Psychometric scale surveys (also known as Likert scale surveys) are commonplace in assessing the perceived importance that the students give to a particular item of the questionnaire \citep[e.g.][]{UnivExtremadura2018,UnivColombia2017,Bernstein2012}.

In particular, this questionnaire summarized in table~\ref{tab:survey} aims to evaluate the degree of achievement that the students perceive related to the abilities described in table~\ref{tab:abilities}. Namely, Q5 in table~\ref{tab:survey} tries to evaluate the degree of achievement of ability 1 in table~\ref{tab:abilities}, Q10 and Q12 are related to ability 2, Q11 is related to ability 3, Q14 to ability 4, and Q3 and Q8 are related to ability 5 in table~\ref{tab:abilities} (some of the statements have been posed in negative form, to try to avoid the acquiescence bias). The rest of questions in table~\ref{tab:survey} evaluate the structure and organization of the workshop and the practical lectures, or are in general broader questions that try to assess whether the students perceive this sort of active learning strategies (and, more specifically, this workshop in particular) as a positive experience overall, with a significant impact in the context of their whole academic training.

It is worthwhile highlighting a few conclusions drawn from the aforementioned survey:

\begin{table}[!t]
\begin{center}
\caption{Answers to the survey conducted among the students enrolled in the \emph{Rocket Engines and Space Propulsion} course (averaged). Q1--Q14 are represented on a 1 to 5 Likert scale (being 1~=~``completely disagree'' and 5~=~``completely agree'' with the statement).} %%% and Q15 provides the total number of hours (averaged)
\label{tab:survey}
\begin{tabularx}{.95\linewidth}{p{0.8\linewidth} *{1}{Y}}
\toprule[1.5pt]\addlinespace[1.5pt]\midrule
\textbf{Q1}. Having taken part in the workshop on the usage of EcosimPro\supers\textregistered~|~\gls{lpres} has significantly contributed to the overall quality of my academic training. & 4.8 \\
\textbf{Q2}. Knowing the usage of EcosimPro\supers\textregistered~|~\gls{lpres} will prove useful for pursuing my future career in the space sector. & 3.9 \\
\textbf{Q3}. The use of simple models to simulate complex systems has helped me to understand the operation and performances of liquid-propellant rocket engines. & 4.3 \\
\textbf{Q4}. Learning computational tools relevant to the space sector is beneficial for expanding and developing the set of skills gained during my academic training. & 4.7 \\
\textbf{Q5}. The guided tutorials in the computer lab have been useful for their intended purpose of teaching me the usage of EcosimPro\supers\textregistered~|~\gls{lpres}. & 4.3 \\
\textbf{Q6}. Learning EcosimPro\supers\textregistered~|~\gls{lpres} has proven too difficult in the short amount of time provided by this course. & 3.2 \\
\textbf{Q7}. The pace of the guided tutorials in the computer lab was too quick, making these practical lectures difficult to follow. & 3.2 \\
\textbf{Q8}. The models chosen for the development of the guided tutorials in the computer lab were interesting to me, and significantly helped me to understand the performances of essential sub-systems of the liquid-propellant rocket engines. & 4.4 \\
\textbf{Q9}. The guided tutorials in the computer lab were posed in such a way that made these practical lectures dynamic and engaging for the students. & 4.0 \\
\textbf{Q10}. The way that the professor guided us during the practical lectures in the computer lab, building up models of gradually increasing complexity, has helped me in grasping the fundamentals of modeling with EcosimPro\supers\textregistered~|~\gls{lpres}. & 4.6 \\
\textbf{Q11}. I have learnt to use EcosimPro\supers\textregistered~|~\gls{lpres} thoroughly enough so as to be able to make use of the aforementioned computational tools on my own. & 3.8 \\
\textbf{Q12}. The number of sessions in the computer lab were just too few, and would have preferred a longer, more in-depth course on the matter, to delve deeper into the details of modeling with EcosimPro\supers\textregistered~|~\gls{lpres}. & 3.8 \\
\textbf{Q13}. The number of sessions in the computer lab were just enough so as to be able to tackle the proposed assignment about the off-design control and performances of liquid-propellant rocket engines on my own, without excessive difficulty. & 4.0 \\
\textbf{Q14}. The proposed assignment was challenging and interesting, and has encouraged me to investigate on my own aspects of the functioning of liquid-propellant rocket engines that have helped me in further understanding their control and performances. & 3.9 \\
%%%\textbf{Q15}. Total dedication, in hours, to the proposed assignment, including the time spent in carrying out the simulations, post-processing the results, and elaborating the final report. & 8.5 \\
\midrule\addlinespace[1.5pt]\bottomrule[1.5pt]
\end{tabularx}
\end{center}
\end{table}

\begin{itemize}
    \item[\checkmark] The students highly value the teaching of computational tools relevant to the space sector as having a very positive impact on their academic training (see Q1 and Q4 in table~\ref{tab:survey}), also benefiting their future professional career (see Q2 in table~\ref{tab:survey}).
    
    \item[\checkmark] It is shown that there is a fairly common agreement among the students in considering the teaching philosophy of \gls{lpres}, using simple models to simulate complex systems, as quite useful to be able to better understand both the operation and performances of the whole system of a liquid-propellant rocket engine, as well as the behavior of its isolated sub-systems (see Q3 and Q8 in table~\ref{tab:survey}).

    \item[\checkmark] The students consider that the practical lectures in the computer lab have indeed accomplished their intended purpose of teaching the fundamentals of modeling with \gls{lpres} (see Q5 and Q10 in table~\ref{tab:survey}), although they are a bit more hesitant as to whether they could make use of the aforementioned tools without having the assistance from the professor (see Q11 in table~\ref{tab:survey}).
\end{itemize}

In addition to the above, the final report that the students must submit on the proposed assignment they were given is used by the professors to evaluate the level of realization that the students have achieved, not only of the abilities gathered in table~\ref{tab:abilities}, but also of the competencies described in table~\ref{tab:competencies}. The distribution of grades obtained by the students in this assignment, based upon that submitted report, is summarized in table~\ref{tab:marks}. These grades showcase that the vast majority of the participants in this modeling workshop did indeed achieve the expected levels of understanding and satisfactorily fulfilled all the evaluated learning objectives, with only 3.7\%{} of the students (excluding those who did not submit) not reaching the minimum score of 5.0 necessary to pass.

\begin{table}[!t]
\begin{center}
\caption{Distribution of grades obtained by the students attending the workshop in their final submitted report.}
\label{tab:marks}
\begin{tabularx}{.6\linewidth}{p{0.35\linewidth} *{1}{Y}}
\toprule[1.5pt]\addlinespace[1.5pt]\midrule
Students’ grade (on a /10 scale) &   \%{} of students            \\ \cmidrule(lr){1-2}
Did not submit	                 &                   15.6 \%{}   \\
Unsatisfactory (4.9 and under)   &   \mbox{\phantom{0}3.1} \%{}  \\
Average (between 5.0 and 7.4)    &                   43.8 \%{}   \\
Good (between 7.5 and 8.9)       &                   25.0 \%{}   \\
Outstanding (9.0 and over)       &                   12.5 \%{}   \\
\midrule\addlinespace[1.5pt]\bottomrule[1.5pt]
\end{tabularx}
\end{center}
\end{table}

For the purposes of calculating the overall grade of the course for the students, the marks they obtained in this assignment were averaged with the marks obtained in a series of other practical assignments (all targeting towards a more \gls{pbl} approach) developed during the whole semester on other topics covered in the syllabus of the course, to account for as much as 40\%{} of their overall grades (the remaining 60\%{} corresponding to the final written exam). In 86.7\%{} of the cases, the overall grades of the students were improved by the marks obtained in this series of practical assignments proposed during the whole semester, which also supports the idea that this sort of complementary workshops and practical activities greatly help the students in understanding the different concepts and acquiring the different learning objectives, and thus improve their overall development in the whole academic program. %%% remove completely this whole paragraph???

%%%%%%%%%%%%%%%%%%%%%%%%%%%%%%%%%%%%%%%%%%%%%%%%%%%%%%%%%%%%%%%%%%%%%%

\subsection{Lessons Learned}\label{sec:results:guidelines}

The development of the \gls{lpres} library is an ongoing process and, as such, the features that we demand from the tool and the requirements that we have imposed are ever-changing as well. Nevertheless, the current state of maturity of the library has already allowed us to establish a core set of the most critical ---or `desirable'--- features that an educational tool should have in order to be suitable for implementing active learning strategies in the engineering classroom. The list provided herein comes from a thorough review of the public literature, as well as by our own experience in this development process and subsequent usage of the library, and can be used either as a checklist to assess the suitability of any readily available computational tool when considering its usage within the engineering classroom, or as a list of requirements to guide the development of new computational tools, irrelevantly of the field considered:

\begin{itemize}
    \item[\textbullet] The tool must have a `professional look', aligned with current standard practices within the industry, which reinforces the perception from the students of acquiring \emph{technological abilities} meaningful for their future professional careers, as shown by the students' responses to the survey in table~\ref{tab:survey}.

    \item[\textbullet] Although the instructor should be in charge of leading and coordinating the different steps of the development process (so that they become \emph{experts} in the usage of the tool, thus enabling \gls{jitt} strategies to be implemented successfully), the coding \emph{per se} should be carried out by the students. The concept of a tool developed \emph{by the students, for the students} increases the involvement of the classroom in the subject, and triggers greater interest to delve deeper into learning \emph{modeling abilities}, as demonstrated by the unprecedented interest exhibited by the students in willing to engage in long-term projects with the Department to be able to contribute to the development of the tool.

    \item[\textbullet] Whenever possible, the development of the tool as a part of an already established modeling environment (and one that is considered within the industry as the current standard practices), allows for long-term collaboration between the industry and the university in the concept of \emph{cooperative education}, greatly benefiting the learners.

    \item[\textbullet] The general philosophy of using ``simplified models'' to simulate ``complex systems'' has proven successful in facilitating the students' better understanding of the phenomena governing the operation and \emph{system dynamics} of rocket engines.

    \item[\textbullet] Additionally, the fact that the code is made available for the learners allows them to see and, far more importantly, to modify the relations between the variables in a given model (including effects or considering new modeling strategies that they might think are relevant for the phenomena in study), providing them with a thorough comprehension of the models' limitations and range of applicability.

    \item[\textbullet] The tool should provide the learners with the means to intuitively model any system they may devise, and be flexible enough to cope with the mathematical burden of the solving process, so that the learners can focus on improving their \emph{systems thinking} and \emph{design thinking} competencies in the context of implementing \gls{pbl} strategies in the classroom.

    \item[\textbullet] Robustness is paramount, both from the numerical stability standpoint as well as by ensuring that the tool yields results with a sufficient degree of fidelity when compared against real-life rocket engines performance data. This provides for the students a solid basis to understand the importance of making quick \emph{estimates} and dealing with \emph{uncertainties} in any engineering design process.
\end{itemize}

%%%%%%%%%%%%%%%%%%%%%%%%%%%%%%%%%%%%%%%%%%%%%%%%%%%%%%%%%%%%%%%%%%%%%%
%%%%%%%%%%%%%%%%%%%%%%%%%%%%%%%%%%%%%%%%%%%%%%%%%%%%%%%%%%%%%%%%%%%%%%
%%%%%%%%%%%%%%%%%%%%%%%%%%%%%%%%%%%%%%%%%%%%%%%%%%%%%%%%%%%%%%%%%%%%%%

\section{Conclusions}\label{sec:conclusions}

The first phase of the project concluded by developing a professional-looking tool, the \gls{lpres} library for EcosimPro\supers{\textregistered}, that has the same operating capabilities as the \gls{espss} professional libraries, but incorporating simple models such as those used to solve practical exercises in the classroom with ``pen and paper''. In consequence, the learning curve for the \gls{lpres} library is shorter, making it possible to do several engine models and calculations in the time and space available in just a few class sessions within the course. The \gls{lpres} library can be used to simulate systems with many elements, which would otherwise be tedious and time-consuming to solve ``by hand'', thus alleviating the `cognitive load' for the learners, who may instead focus on the development of their \emph{systems thinking} and \emph{design thinking} competencies.

The most important features for a software tool to have in the context of implementing active learning strategies in the engineering classroom were summarized and discussed in Section~\ref{sec:results}.

Despite the fact that not every student managed to become absolutely confident in the usage of the tool, most of them acknowledged to be autonomous in the tasks of modeling and simulating, with only minor issues ---typically related to numerical convergence problems (which many times are straightforward to solve for a more experienced user) and not with the modeling itself--- occurring from time to time, and hence the two important conclusions from this study remain valid:
\begin{inparaenum}[\itshape a\upshape)]
    \item students' perception of the quality of their training improved, they felt more motivated and engaged in the subject more willingly, as shown by the survey carried out; and
    \item their grade of retention and understanding of the matter increased very significantly, as revealed by the pre- and the post-test scores achieved by the students, as well as by the degree of acquaintance with the learning objectives evaluated with their final submitted report.
\end{inparaenum}

Given the positive impact observed on the learners, both in their understanding and in their involvement within the subject of \emph{Rocket Engines and Space Propulsion}, the practicum has been definitely incorporated into the standard program for the course, and is nowadays considered an essential part of the curriculum.

We expect that the results presented will encourage instructors in other fields with similar challenges for implementing active learning strategies to develop their own computer tools, and we firmly believe that the list provided of required features identified for such a tool to meet the demands of its intended usage can be made extensible to any other discipline, surpassing just the engineering (and, more precisely, the rocket science) scope, as reported in this paper.

%%%%%%%%%%%%%%%%%%%%%%%%%%%%%%%%%%%%%%%%%%%%%%%%%%%%%%%%%%%%%%%%%%%%%%
%%%%%%%%%%%%%%%%%%%%%%%%%%%%%%%%%%%%%%%%%%%%%%%%%%%%%%%%%%%%%%%%%%%%%%
%%%%%%%%%%%%%%%%%%%%%%%%%%%%%%%%%%%%%%%%%%%%%%%%%%%%%%%%%%%%%%%%%%%%%%

\glsresetall   % re-explain acronyms as if they appeared for the first time
\section*{Acknowledgements}

We would like to thank the \gls{upm}, Spain, for funding this project, as part of the 2014 edition of the \emph{``Ayudas a la Innovación Educativa y a la Mejora de la Calidad de la Enseñanza''} program; and Empresarios Agrupados, for the long-lasting collaboration that we have established over the last several years, as well as for their continuous support in the usage of their tools.   % for the double-blind review, HIDE THIS !!!

%%%%%%%%%%%%%%%%%%%%%%%%%%%%%%%%%%%%%%%%%%%%%%%%%%%%%%%%%%%%%%%%%%%%%%
%%%%%%%%%%%%%%%%%%%%%%%%%%%%%%%%%%%%%%%%%%%%%%%%%%%%%%%%%%%%%%%%%%%%%%
%%%%%%%%%%%%%%%%%%%%%%%%%%%%%%%%%%%%%%%%%%%%%%%%%%%%%%%%%%%%%%%%%%%%%%

\bibliography{main}

%%%%%%%%%%%%%%%%%%%%%%%%%%%%%%%%%%%%%%%%%%%%%%%%%%%%%%%%%%%%%%%%%%%%%%
%%%%%%%%%%%%%%%%%%%%%%%%%%%%%%%%%%%%%%%%%%%%%%%%%%%%%%%%%%%%%%%%%%%%%%
%%%%%%%%%%%%%%%%%%%%%%%%%%%%%%%%%%%%%%%%%%%%%%%%%%%%%%%%%%%%%%%%%%%%%%

\end{document}